\documentclass{aa}
\usepackage{graphicx}
\usepackage{amsmath}
\RequirePackage{natbib}
\begin{document}
\title{Acoustic wave propagation in the solar sub-photosphere with localised 
magnetic field concentration: effect of magnetic tension}
\author{S. Shelyag, S. Zharkov, V. Fedun, R. Erd\'{e}lyi, M.J. Thompson}
\titlerunning{Acoustic wave propagation through magnetic field concentration}
\authorrunning{Shelyag et al.}
\institute{Solar Physics and Space Plasma Research Center, Department 
of Applied Mathematics, University of Sheffield, Hicks Building,
Hounsfield Rd., Sheffield, S7 3RH, United Kingdom}
\date{01.01.01/01.01.01}
\abstract{}
{In this paper we analyse numerically the propagation and dispersion of acoustic 
waves in the solar-like sub-photosphere with localised non-uniform magnetic field
concentrations, mimicking sunspots with various representative magnetic field configurations.}
{Numerical simulations of wave propagation through the solar sub-photosphere with
a localised magnetic field concentration are carried out using SAC, which solves 
the MHD equations for gravitationally stratified plasma. 
The initial equilibrium density and pressure stratifications are derived from 
a standard solar model. Acoustic waves are generated by a source located at the 
height approximately corresponding to the visible surface of the Sun. We analyse
the response of vertical velocity to changes in the interior 
due to magnetic field at the level corresponding to the visible 
solar surface, by the means of local time-distance helioseismology.}
{The results of numerical simulations of acoustic wave propagation and 
dispersion in the solar sub-photosphere with localised magnetic field 
concentrations of various types are presented.  Time-distance diagrams of the vertical 
velocity perturbation at the level corresponding to the visible solar 
surface show that the magnetic field perturbs and scatters acoustic waves 
and absorbs the acoustic power of the wave packet. For the weakly magnetised case
the effect of magnetic field is mainly thermodynamic, since the magnetic field
changes the temperature stratification. However, we observe the signature of 
slow magnetoacoustic mode, propagating downwards, for the strong magnetic field cases.
}
{}

\maketitle

\section{Introduction}

The internal structure of sunspots is still not well known. 
Helioseismological
techniques, which analyse the influence
of internal solar inhomogeneities on sound wave propagation 
and the signatures of such waves at the solar surface, might 
be of great help in revealing the invisible, sub-photospheric solar processes.
The ability of forward numerical simulations to predict and model a number of
solar phenomena in helioseismology has been shown by e.g. \citet{shelyag4,shelyag2,
hanasoge1,parchevskii1} and others. As magnetic fields are, perhaps, the
most important property of many solar features, a new and  rapidly developing 
field is the study of the influence of magnetic fields on acoustic wave propagation of solar magnetic 
field concentrations such as sunspots or solar active regions. 
The appearance and importance of 
slow magnetoacoustic waves has been shown in forward MHD simulations in  
polytropic models by \citet{cally1, gordovskyy, moradi1}. Ray-approximation simulations in 
a more realistic and applicable magnetised model have shown a similar behaviour to the acoustic waves 
\citep{cally2}. \citet{shelyag2} have investigated the influence of sub-photospheric 
flows on acoustic wave propagation using foward modelling and demonstrated a discrepancy
between the actual flow profiles and the flow profiles obtained by ray-approximation inversion. 
The simulations of a wave packet, constructed from $f$-modes, which was carried out by \citet{cameron},
have shown a good agreement with helioseismological observations of sunspots. Now it is timely to 
perform a full forward magneto-hydrodynamic simulation of a wave packet 
propagating through a non-uniform magnetic field region in the solar sub-photosphere with
a realistic temperature profile. 

In the simulations presented here we consider three different representative configurations 
of solar magnetic field. Each of the cases has a common feature that the field is spatially localised, 
allowing a direct comparison of the
travel speeds and time differences of the wave propagation between the magnetised
and non-magnetised solar plasma. The representative configurations
differ in both magnetic field strength and geometry. Their spatial structure 
affects the temperature stratification 
of the simulated sunspot by the magnetic tension. We selected two representative strong 
field configurations
with opposite effects on the temperature in the sunspot: one, where the magnetic-field 
curvature is strong and, thus, increases the temperature in the magnetic field
region; and another, where the magnetic field curvature is small, and the temperature 
is decreased in the sunspot. The magnetic configurations
we apply are in magnetohydrostatic equilibrium with the ambient external plasma.
These two-dimensional magnetic fields are self-similar and non-potential.

We present the results according to the following general structure.
Section 2 briefly describes the numerical techniques that we have used to carry out the simulations.
The configurations of the magnetic fields and initial configuration for the simulations are
described in Section 3. The source used to generate the acoustic modes in the numerical domain 
is presented in Section 4. Section 5 is devoted to (i) the techniques of helioseismological analysis 
we have used, and (ii) the results we obtained. Section 6 concludes.

\section{Simulation model}

The code SAC (Sheffield Advanced Code) has been developed by \citet{shelyag1} 
to carry out numerical studies. The code is based on VAC \citep[Versatile Advection Code,][]{toth1998},
however, it employs artificial diffusivity and resistivity in order to stabilise
the numerical solutions. Also, SAC uses the technique of variable separation to background and perturbed 
components to treat gravitationally stratified plasma. 
According to \citet{shelyag1}, if a plasma is assumed to be in magnetohydrostatic equilibrium given by
\begin{equation}
\left(\mathbf{B}_b \cdot \nabla \right) \mathbf{B}_b + \nabla\frac{\mathbf{B}^2_b}{2}+\nabla p_b=\rho_b
\mathbf{g},
\label{eq:mhdeq0}
\end{equation}
the system of MHD equations governing arbitrary perturbations of density, momentum, energy and 
magnetic field is written as follows:
\begin{equation}
\frac{\partial \tilde{\rho}}{\partial t} + \nabla \cdot \left[{\bf{v}} 
\left(\rho_b+\tilde{\rho}\right) \right]=0+D_{\rho}\left(\tilde{\rho} \right),
\label{eq:mhdeq1}
\end{equation}
\begin{equation}
\begin{split}
\frac{\partial{\left[\left(\rho_b+\tilde{\rho}\right)\bf{v}\right]}}{\partial t}+
\nabla\cdot\left[\bf{v}\left(\rho_b+\tilde{\rho}\right)
\bf{v}-\tilde{\bf{B}}\tilde{\bf{B}}\right]- \\
\nabla\cdot\left[\tilde{\bf{B}}{\bf{B}}_b+{\bf{B}}_b\tilde{\bf{B}}\right]
+\nabla {\tilde{p}_{t}}=\tilde{\rho} {\bf{g}} + {\bf{D}}_{\rho v}\left[\left(\tilde{\rho}+
\rho_b \right){\bf{v}}\right],
\end{split}
\label{eq:mhdeq2}
\end{equation} 
\begin{equation}
\begin{split}
\frac{\partial{\tilde{e}}}{\partial{t}}+\nabla \cdot \left[ {\bf{v}}\left(e+e_b\right)-
{\tilde{\bf{B}}\tilde{\bf{B}}\cdot\bf{v}}+{\bf{v}}\tilde{p}_{t}\right]-\\
\nabla\cdot\left[\left(\tilde{\bf{B}}{\bf{B}}_b+{\bf{B}}_b\tilde{\bf{B}}\right) 
\cdot {\bf{v}} \right]+p_{tb}\nabla{\bf{v}}-{{\bf{B}}_b{\bf{B}}_b}\nabla{\bf{v}}= \\
=\tilde{\rho} {\bf{g}} \cdot {\bf{v}} +D_e\left(\tilde{e} \right),
\end{split}
\label{eq:mhdeq3}
\end{equation}
\begin{equation}
\frac{\partial{\tilde{\bf{B}}}}{\partial{t}}+\nabla\cdot\left[{\bf{v}}
(\tilde{\bf{B}}+{\bf{B}}_b)-(\tilde{\bf{B}}+{\bf{B}}_b){\bf{v}}\right]=0 + {\bf{D}}_{B}\left(\tilde{\bf{B}}\right),
\label{eq:mhdeq4}
\end{equation}
where $\tilde\rho$ and $\rho_b$ are the perturbation and background density counterparts, 
${\bf{v}}$ is the total velocity vector, $e$ is the total background energy density per unit 
volume, $\tilde e$ is the perturbed energy density per unit volume, ${\bf{B}}_b$ and 
$\tilde{\bf{B}}$ are the background and perturbed magnetic field vectors, $p_{tb}$ is the total 
(magnetic $+$ kinetic) background pressure, $\gamma$ is the 
adiabatic gas index, ${\bf{g}}$ is the external gravitational field vector, and $\tilde{p}_t$ 
is the perturbation to the total pressure
\begin{equation}
\tilde{p}_t=\tilde{p}_{k}+\frac{{\tilde{\bf{B}}}^2}{2}+{\bf{B}}_b{\tilde{\bf{B}}},
\label{eq:mhdeq6}
\end{equation}
or, in terms of perturbed energy density per unit volume $\tilde{e}$,
\begin{equation}
\tilde{p}_{k}=\left(\gamma-1\right)\left(\tilde{e}-\frac{\left(\rho_b+\tilde{\rho}\right){\bf{v}}}{2}-
{\bf{B}}_b {\tilde{\bf{B}}}-\frac{{\tilde{\bf{B}}}^2}{2}\right),
\label{eq:mhdeq6b}
\end{equation}
and
\begin{equation}
\begin{split}
\tilde{p}_t=\left(\gamma-1\right)\left[\tilde{e}-\frac{\left(\rho_b+\tilde{\rho}\right){\bf{v}}^2}{2}\right]- \\
- \left(\gamma-2\right)\left({\bf{B}}_b{\tilde{\bf{B}}}+\frac{{\tilde{\bf{B}}}^2}{2}\right).
\end{split}
\label{eq:mhdeq6a}
\end{equation}
Here $p_{tb}$ denotes the total background pressure
\begin{equation}
p_{tb}=p_{kb}+\frac{{\bf{B}}_b^2}{2}
\label{eq:mhdeq7}
\end{equation}
which, in terms of background conservative variables, gives
\begin{equation}
p_{kb}=\left(\gamma-1\right)\left(e_b-\frac{{{\bf{B}}}_b^2}{2}\right),
\label{eq:mhdeq7b}
\end{equation}
and
\begin{equation}
p_{tb}=\left(\gamma-1\right) e_b - \left(\gamma-2\right)\frac{{{\bf{B}}}_b^2}{2}.
\label{eq:mhdeq7a}
\end{equation}

\begin{figure}
\includegraphics[width=1.0\linewidth]{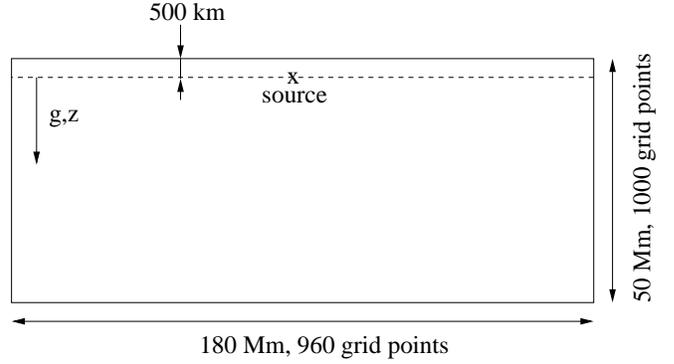}
\caption{Sketch of the simulation domain geometry, used in the simulations.}
\label{fig:d1}
\end{figure}

Eqs.~(\ref{eq:mhdeq1})-(\ref{eq:mhdeq7a}) are solved using a fourth-order central difference
scheme for the spatial derivatives and are advanced in time by implementing a fourth order
Runge-Kutta numerical method. The source terms $D$ in the right-hand sides of the
equations denote the artificial diffusivity and resistivity terms.
 The simulation domain is shown in Fig.~\ref{fig:d1}. The 2D box is $180~\mathrm{Mm}$ wide 
and $50~\mathrm{Mm}$ deep, and has a resolution of 960x1000 grid points; the upper boundary 
of the domain is at the solar surface $R=R_{\odot}$. The boundaries of the domain 
are open. The perturbation source is located in the upper-middle ($500~\mathrm{km}$ below the 
upper boundary) of the simulation box. The synthetic measurement level is located at the
solar surface.

\section{Magnetic fields and initial conditions}

For an initial background model, we adopt the Standard Model S \citep{cdmodel}. The model
is then adjusted to have the same temperature stratification as the Standard Model, if the
constant adiabatic constant $\Gamma_1$ is taken.
According to the Standard Model S, the pressure at the solar surface
$R=R_{\odot}$ is equal to $p_{\odot}=7.61 \cdot 10^4~\mathrm{dyn/cm^2}$. This creates
an upper limit for the vertical straight uniform magnetic field in the magnetohydrostatic
equilibrium with non-magnetic external plasma to be about 
$B_{max}=\sqrt{8\pi p_{\odot}}=1.4~\mathrm{kG}$. The measured magnetic field strength
in sunspot umbrae is about $2.5-3.5~\mathrm{kG}$, which suggests the implementation of curved magnetic fields
in the simulations. In the case of curved magnetic field, magnetic tension
balances magnetic pressure, thus increasing the upper limit for the equilibrium
magnetic field. An example of such balance is a potential magnetic field, where
the magnetic tension is exactly equal to the magnetic pressure: such a field does 
not change background pressure and
density equilibrium. However, the potential magnetic fields have a disadvantage
for numerical modelling. If a potential magnetic field is considered, the boundaries
of the numerical domain should be either fixed or periodic to confine the magnetic field
and prevent it from strong expansion. 

We use a self-similar non-potential magnetic field configuration 
\citep{schltem, schremp, cameron}, which can be obtained from the following set of equations:
\begin{equation}
B_x=-\frac{\partial f}{\partial z} \cdot G\left(f\right),
\label{mf:eq1}
\end{equation}
\begin{equation}
B_z=\frac{\partial f}{\partial x} \cdot G\left(f\right),
\label{mf:eq2}
\end{equation}
and
\begin{equation}
f=x \cdot B_{0z}\left(z\right),
\label{mf:eq3}
\end{equation}
where $B_{0z}$ describes the decrease of the vertical component of magnetic field towards 
the top of the model, and $G$ is the function which defines how the magnetic field opens 
up with height. The magnetic field constructed in this way is divergence-free by definition. 
The equilibrium background gas pressure and density are then recalculated using the 
magnetohydrostatic equilibrium 
condition Eq.~(\ref{eq:mhdeq0}). If the magnetic field $\mathbf{B}_b$ is prescribed,
Eq.~(\ref{eq:mhdeq0}) splits into two independent equations for the pressure and
density deviations from the initial state, caused by the magnetic field. These
equations are then solved numerically in order to obtain the gravitationally stratified 
plasma model with localised magnetic field concentration in magnetohydrostatic equilibrium.

\begin{figure}
\includegraphics[width=1.0\linewidth]{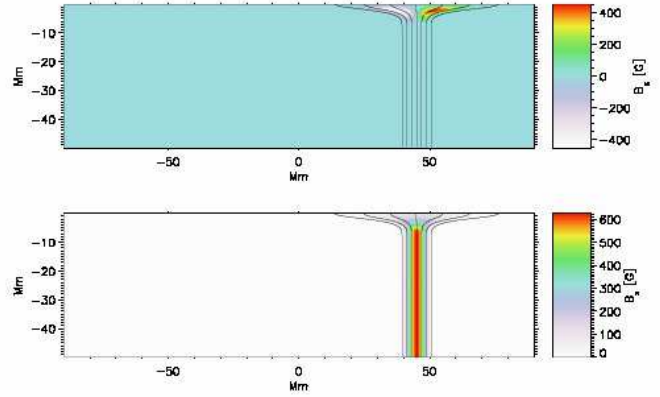}
\caption{Magnetic field configuration for the case of weak magnetic field (Case A) in the 
sub-photospheric domain of the size of 50 Mm in vertical and 180 Mm in horizontal direction. The 
horizontal ($B_x$) and vertical ($B_z$) components of the magnetic field are shown. The field 
lines are overplotted.}
\label{mf:f1}
\end{figure}

\begin{figure}
\includegraphics[width=1.0\linewidth]{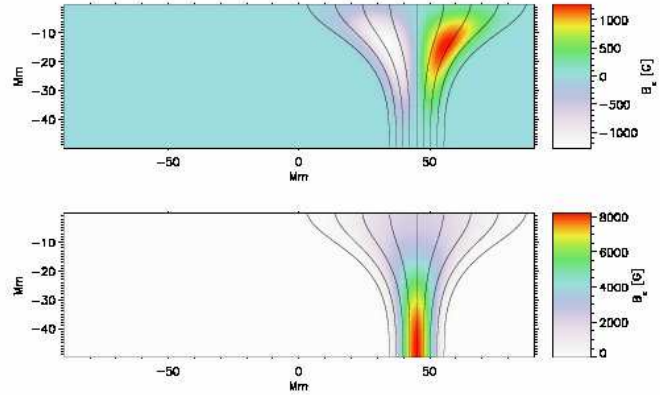}
\caption{Same as Fig.~\ref{mf:f1}, with a strong magnetic flux 
($B_{z,\odot}=3.5~\mathrm{kG}$) but with weakly-curved magnetic field lines, Case B.}
\label{mf:f2}
\end{figure}

\begin{figure}
\includegraphics[width=1.0\linewidth]{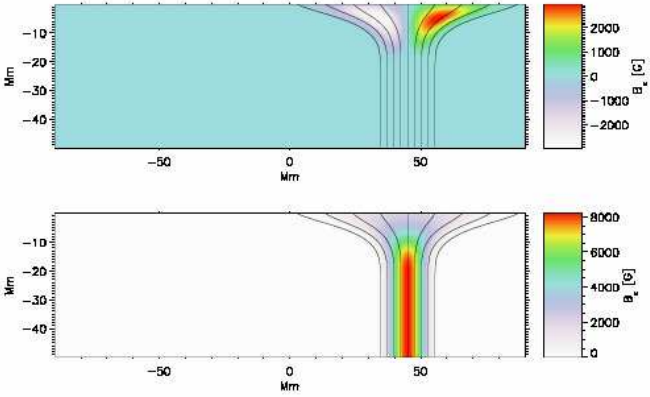}
\caption{Same as Fig.~\ref{mf:f1}, with a strongly-curved, strong ($B_{z,\odot}=3.5~\mathrm{kG}$) 
magnetic field, Case C.}
\label{mf:f3}
\end{figure}

Three characteristic situations, mimicking sunspots which differ by the magnetic 
field strength at the visible solar surface and curvature of the magnetic field, 
are chosen for helioseismic analysis: weak but strongly-curved 
magnetic field ($B_{z,\odot}=120~\mathrm{G}$, Case A), strong but weakly-curved magnetic field 
with $B_{z,\odot}=3.5~\mathrm{kG}$ (Case B), and strongly-curved strong magnetic field 
($B_{z,\odot}=3.5~\mathrm{kG}$, Case C). The magnetic field structures for these situations
are shown in Figs.~\ref{mf:f1}-\ref{mf:f3}.

\begin{figure}
\centering
\includegraphics[width=1.0\linewidth]{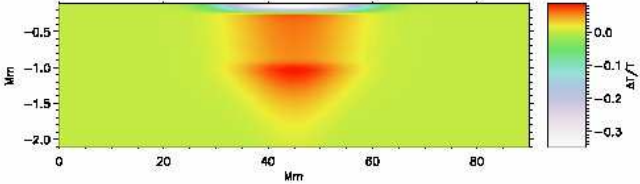}
\caption{Zoom-in of the temperature difference in the magnetic field region for Case A. 
The temperature increase is caused by magnetic tension.}
\label{mf:t1}
\end{figure}

The curvature of magnetic field changes the temperature stratification in the domain.
For the first case of the weak magnetic field (Case A, Fig.~\ref{mf:f1}), the temperature 
change is rather small (Fig.~\ref{mf:t1}). Below the temperature decrease, which is caused 
by the drop of the kinetic pressure in the region where the magnetic field is nearly 
vertical, a temperature increase is noticeable. This increase is caused by the 
pressure rise which is needed to compensate the increase in magnetic tension.

\begin{figure}
\centering
\includegraphics[width=1.0\linewidth]{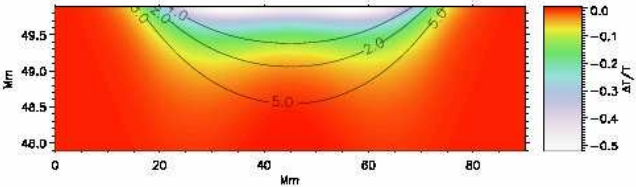}
\caption{Same as Fig.~\ref{mf:t1}, with a strong magnetic flux and weakly-curved 
magnetic field lines, Case B.
Here, the temperature decreases at the solar surface. This temperature decrease is 
caused by magnetic pressure.
Plasma $\beta$ contours with levels $\beta=1,~2,~5$ are overplotted.
}
\label{mf:t2}
\end{figure}

\begin{figure}
\centering
\includegraphics[width=1.0\linewidth]{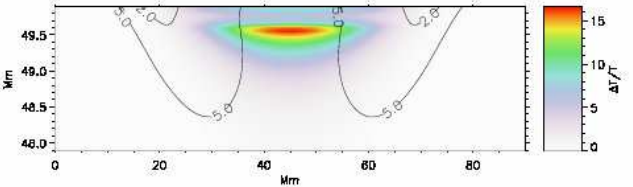}
\caption{Same as Fig.~\ref{mf:t1}, with the strongly-curved and strong magnetic field, Case C.
In this case magnetic tension prevents evacuation of the magnetic region, and the temperature
is increased.}
\label{mf:t3}
\end{figure}

The two initial configurations with strong magnetic field show these processes in greater 
detail. For Case B (weakly-curved magnetic field, Fig.~\ref{mf:f1}), the temperature
is considerably decreased beneath the solar surface (Fig.~\ref{mf:t2}), because the magnetic
field is nearly vertical at the surface. The effect of magnetic tension reveals itself 
in the strongly-curved magnetic field configuration (Case C, Fig.~\ref{mf:t3}). Here, in 
this latter case, the temperature deviation $\Delta T/T$ is mainly positive.

\section{Acoustic source}

To generate acoustic waves, we introduce a perturbation source described by the expression:
\begin{equation}
v_z = A_0 \sin \frac{2 \pi t}{T_0} \exp\left( -\frac{\left(t-T_1\right)^2}{\sigma_1^2}\right) 
\exp \left( -\frac{\left(r-r_0\right)^2}{\sigma_0^2}\right),
\end{equation}
where $T_0$=300 s, $T_1$=600 s, $\sigma_1$=100 s, $\sigma_2$=0.1 Mm, $r_0$ is the source 
location. The source is located in the middle of the horizontal layer slightly beneath
the solar surface (see Fig.~\ref{fig:d1}). The amplitude of the source $A_0$ is chosen
to be sufficiently small, which ensures that convective processes will not be initiated
in the otherwise convectively unstable equilibrium, and that the perturbation is still linear, i.e.
does not change the background strongly.
The source generates a temporaly localised wave packet with the duration of 
about $600~\mathrm{s}$, which has the main frequency of about $3.33~\mathrm{mHz}$.

\begin{figure}
\includegraphics[width=1.0\linewidth]{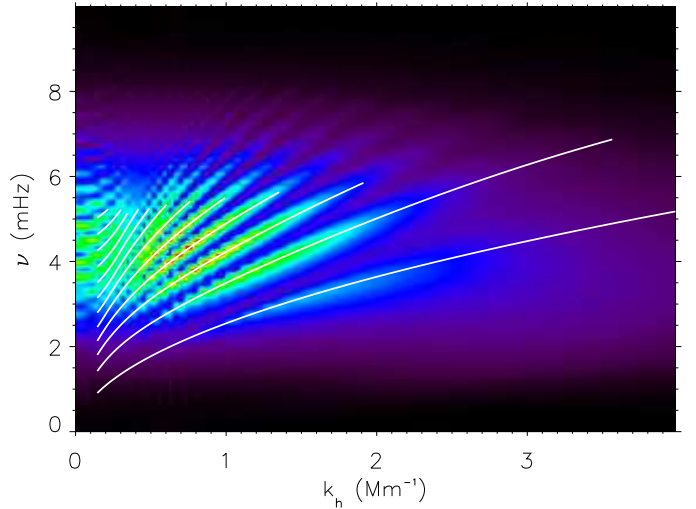}
\caption{Power spectrum of the vertical velocity perturbation  generated by the source. 
The $p$ modes are visible up to high orders. Eigenmodes of the background model are 
overplotted by solid lines.}
\label{as:f1}
\end{figure}

The acoustic response of the simulation box to the source is shown in Fig.~\ref{as:f1}.
It is evident from the figure that the source generates a whole branch of various solar
acoustic modes. The $p$-modes are visible up to high orders. 

Also, in order to check the validity of the simulations, the one-dimensional calculation of the 
eigenmodes of the initial background model was performed. The corresponding eigenfrequencies are  
overplotted in the figure (solid lines). A small difference between 
the model and the observed frequencies is caused by discrepancies between the Standard 
Model S and the model implemented here, which is accountable for using the 
equation of state for an ideal gas. However, we leave the development of non-ideal equation
of state, including the ionisation processes, and 
correction of these discrepancies for a future work, since they will not play any
major role in the calculations of the influence of non-uniform magnetic fields
on acoustic wave propagation in the solar photosphere.

\begin{figure}
\includegraphics[width=1.0\linewidth]{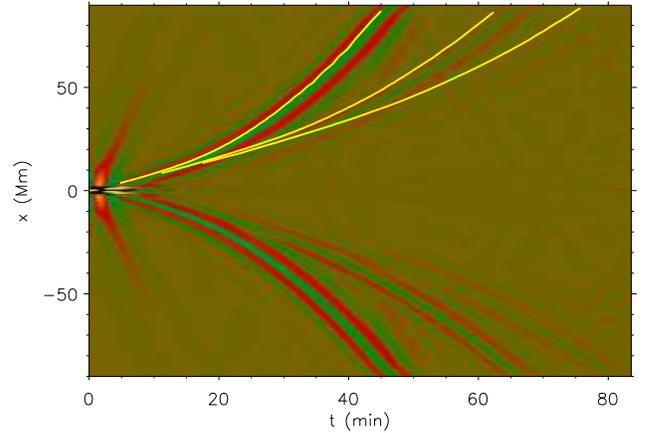}
\caption{Cross-correlation function deduced from 
the vertical velocity perturbation at the solar surface,
generated by the acoustic source. Three wave bounces are clearly visible.  Group travel times 
for the first three bounces deduced from ray theory are overplotted.}
\label{as:f2}
\end{figure}

Next, Fig.~\ref{as:f2} shows the time-distance diagram computed at the simulated solar surface 
via cross-correlating the vertical velocity component, generated by the acoustic source.
Three wave bounces are clearly visible on the plot. Some weak and artificial reflection from the
side boundaries, which is caused by not perfectly transparent boundaries of the 
numerical domain, is also noticeable. 

\section{Time-distance analysis}
It has been mentioned above that the acoustic source is located in the middle of the 
horizontal layer close to the solar surface. This allows us to study the influence 
of the magnetic field on the acoustic response of the simulated solar 
sub-photosphere by comparing the plasma velocities to the left (non-magnetic 
part of the domain) and to the right (where the magnetic field is implemented) 
from the source. Similar technique has already been proposed and used by \citet{shelyag2}
to reveal the discrepancies between the real and inverted velocity profiles 
for the sub-photosphere with embedded sub-photospheric horizontal flows. 

The vertical velocity differences are computed between the points, located at
the same distance to the left and to the right from the source. The difference 
images, obtained in this way, revail the phase shifts and amplitude changes 
the wave packets experience due to the propagation in the magnetised region, 
compared to the non-magnetised one.

In each case of a magnetic configuration we compute the acoustic power of the vertical component of
the velocity oscillations over the period 
of the simulation as function of horizontal and depth coordinates, 
$$a_p(x, z)=\int v_z^2(x, z, t) dt,
$$
and then consider the ratio
between the corresponding points of the quiet Sun and perturbed parts of the model. 
A cut at the surface level corresponds to the acoustic power measurements deduced 
observationally. In addition, we have measured the travel time perturbations by 
cross-correlating the velocity signal at the source location with the signal at 
the target location, taking the quiet Sun cross-correlation function as a reference 
and using both G{\'a}bor wavelet fitting \citep{Kosovichev} or linearised definition 
outlined by \citet{gizonbirch}.

In the sections \ref{ssc1}-\ref{ssc3}  we analyse by these means the wave propagation through 
the three cases of magnetic field structures. In the section \ref{ssc4}, the results are compared. 

\subsection{Weak magnetic field (Case A)}
\label{ssc1}

The analysis shows that for the configuration of a weak magnetic field 
(Fig.~\ref{mf:f1}) the influence 
the magnetic field exercises on the wave propagation is mainly caused by 
temperature (and hence local sound speed) changes in the magnetised region. 
The temperature increase below the simulated sunspot (see Fig.~\ref{mf:t1}) 
causes negative phase shifts of the wave packets propagating through the magnetic structure. 

\begin{figure}
\includegraphics[width=1.0\linewidth]{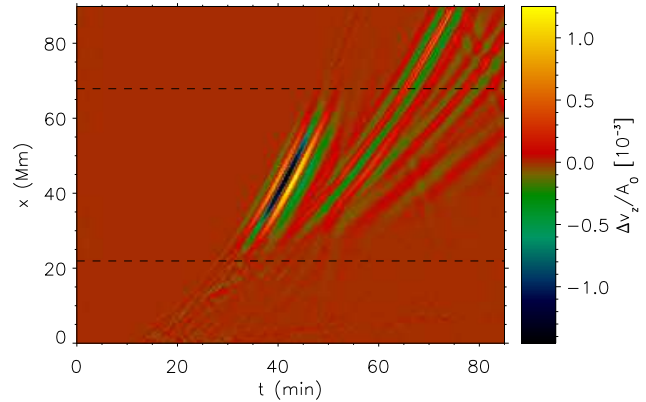}
\caption{Vertical speed difference image for Case A. 
Difference values are computed between points located at the same distance 
but opposite sides of the source. Two dashed lines bound the magnetic region 
with $|B|>25~\mathrm{G}$. The first bounce (leftmost in the figure) is affected 
only locally by the magnetic field, however, the second and third 
bounces are also affected in the 60-80 Mm distance region.}
\label{td:f1}
\end{figure}

In Fig.~\ref{td:f1} small phase 
shifts are observed in all of the bounces, however, the first bounce wave is only 
affected in the magnetic field region (the phase shifts for 
the first bounce at 0-30 Mm and at 70-90 Mm are of the order of numerical noise). 
It was found that higher order bounces are all affected from 30 Mm distance onwards from the source.

\begin{figure}
\includegraphics[width=1.0\linewidth]{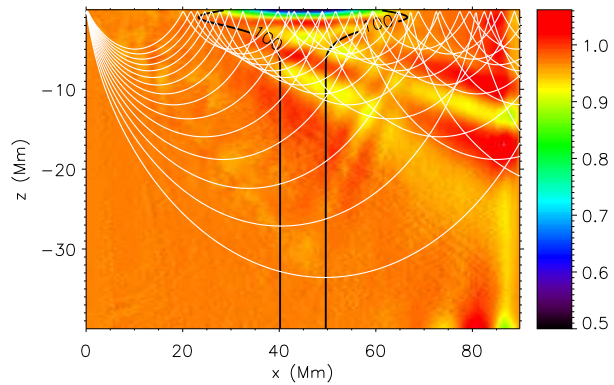}
\caption{Synthetic acoustic power ratio image for the weakly magnetised region of the 
weakly magnetised model (Case A). 
The ratios are computed between the points located at the same distance and opposite sides from 
the source. The image shows the regions of acoustic power decreased compared with the 
ambient non-magnetic medium. The black lines are the contours of vertical magnetic field 
at 50 and $100~\mathrm{G}$, respectively.
The acoustic rays for e.g. the frequency $f=4.5~\mathrm{mHz}$ computed for the quiet Sun model are also 
overplotted. 
}
\label{td:f2}
\end{figure}

In Fig.~\ref{td:f2} we present the acoustic power ratio measured in the vertical velocity
as a function of two spatial coordinates between the waves propagating
in the quiet Sun and in the perturbed part of the model. We found that the power deficit at the
location of the flux tube is confined to the uppermost layers of the model with
the acoustic power ratio falling the closer to the surface one measures at. 
The ratio is constant and equal to unity to the left of the flux tube, while to the right 
we observe variations in the acoustic power at depths of 3 Mm and lower, extending along 
the straight lines starting below the surface at the flux tube boundary. From the overplotted acoustic rays 
computed for the unperturbed model we find that these lines agree well  with the second and third
bounce ray envelopes.

\subsection{Weakly-curved, strong magnetic field (Case B)}
\label{ssc2}
\begin{figure}
\includegraphics[width=1.0\linewidth]{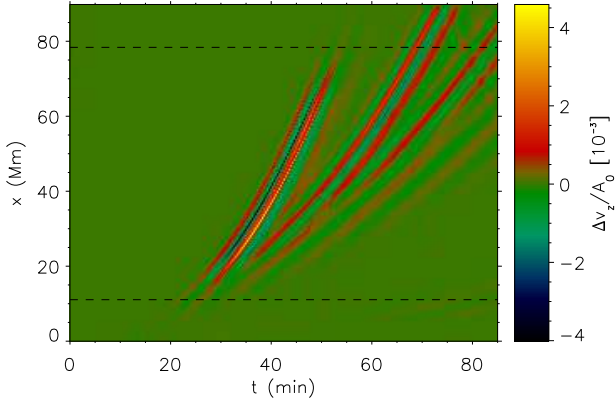}
\caption{Same as Fig.~\ref{td:f1} but for the strong weakly-curved magnetic field 
($B_{z,\odot}=3.5~\mathrm{kG}$), case B. Two dashed lines bound the magnetic region 
with $|B|>250~\mathrm{G}$.}
\label{td:f3}
\end{figure}

In Case B 
(see Fig.~\ref{mf:f2}), the temperature distribution is such that the temperature 
decreases in the sunspot region (Fig.~\ref{mf:t2}), similarly to Case A, 
since the magnetic tension is relatively small. Again, as for Case A,
accordingly to the vertical velocity difference image Fig.~\ref{td:f3}, the first bounce 
is affected by the magnetic field only in the magnetic field region. 
It is found that the second
and higher-order bounces carry information about the interaction with magnetic field also in the 
non-magnetic or weakly magnetised sub-surface regions. 
Intuitively, following the temperature structure, a delay in arrival time of the wave packet at a distance 
from the source is expected, since the sound speed in the simulated sunspot 
is lower than in the non-magnetic surrounding plasma. A more detailed analysis nicely confirms this 
expectation. The first ridge in the image of velocity difference is positive, meaning that
the wave arrives later at the point in the sunspot when compared with the counterpart wave that arrives at
the same distance in the non-magnetic plasma.

\begin{figure}
\includegraphics[width=1.0\linewidth]{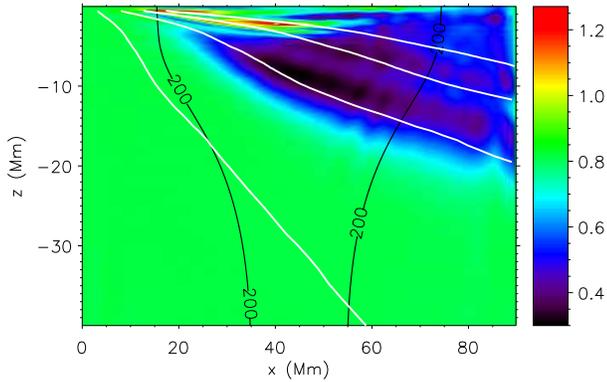}
\caption{Kinetic energy density ratio for the strong magnetic flux but weakly-curved 
magnetic field lines, Case B. 
The lower turning points for the first four bounces of rays emanating from the source 
are overplotted in white color.  The black lines are the contours of the magnetic 
field $|B|=200~\mathrm{G}$.}
\label{td:f4}
\end{figure}

\begin{figure}
\includegraphics[width=1.0\linewidth]{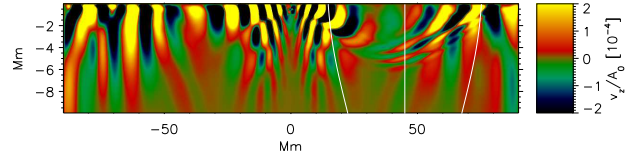}
\caption{Snapshot of the horizontal component of the velocity field in the upper layers of the domain, 
taken at $t=5130~\mathrm{s}$ of the simulation for Case B. The slow mode is visible in the 
magnetised region beneath the solar surface between $x=20~\mathrm{Mm}$ and $x=70~\mathrm{Mm}$.
The magnetic field lines are overplotted.}
\label{td:vx1}
\end{figure}

In Case B, the slow magneto-acoustic mode is also observed in the domain
(see Fig.~\ref{td:vx1}). Generally, the slow-wave motions follow the magnetic field 
structure and shape. Further, a suppression of oscillations is observed at the surface in the 
magnetised region. At the distance of $40~\mathrm{Mm}$ to the right from the source, 
the amplitude ratio of horizontal velocity oscillations at the surface to the source 
amplitude $A_0$ is about
$0.0001$, while at the distance of $-40~\mathrm{Mm}$ to the left from 
the source it is more than $0.0004$ (note that the image is overexposed
in order to reveal the small amplitude structures). Thus, a significant part of
oscillation energy transforms into slow magnetoacoustic wave motion, which propagates
downwards along the magnetic field lines, and is taken out from the surface. 

The kinetic energy density ratio plot (Fig.~\ref{td:f4}) shows the lines of
decreased ratio, similar to Case A, with the structure immediately below the surface 
of the tube showing greater 
complexity than before, perhaps, due to the effect of magnetoacoustic waves.
Fig.~\ref{td:f4} also shows the curves corresponding to lower turning points for the 
first four bounces of the quiet Sun rays. Note that only the second, third and fourth
bounce turning points correspond to the ray envelopes (see Fig.~\ref{td:f2}), representing caustic 
surfaces \citep{kravorl}.

\subsection{Strongly-curved, strong magnetic field (Case C)}
\label{ssc3}
\begin{figure}
\includegraphics[width=1.0\linewidth]{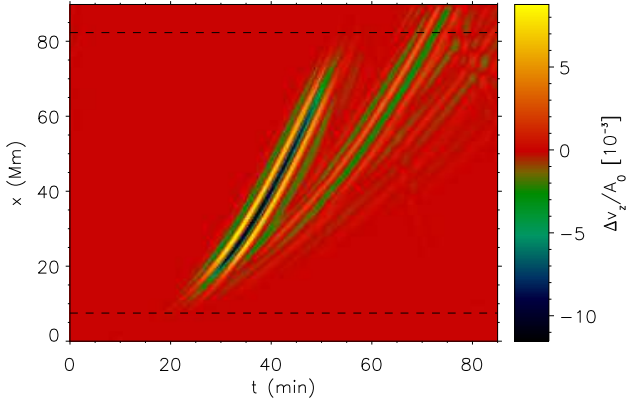}
\caption{Same as Fig.~\ref{td:f1}, but for the strongly-curved, strong magnetic field (Case C).
Two dashed lines bound the magnetic region with $|B|>250\mathrm{G}$.}
\label{td:f5}
\end{figure}

Strong magnetic tension in Case C (Fig.~\ref{mf:f3}) changes 
the kinetic gas pressure in a way that 
the temperature (and hence sound speed) in the sunspot region increases (see Fig.~\ref{mf:t3}).
As expected, the sound speed increase leads to a faster wave propagation through
the magnetised region. Correspondingly, the vertical speed difference image (Fig.~\ref{td:f5}) 
shows a negative sign of the
first ridge in the first bounce. Similar to the previous cases, the first bounce is
affected by magnetic field only in the magnetic region ($20-70~\mathrm{Mm}$ distance),
and the second and higher order bounces are affected everywhere after $20~\mathrm{Mm}$ onwards.

\begin{figure}
\includegraphics[width=1.0\linewidth]{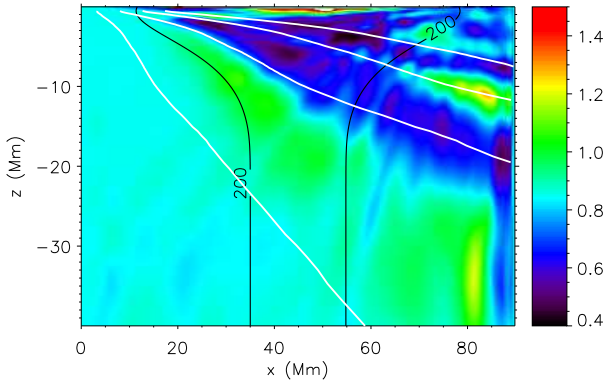}
\caption{Oscillatory power in pressure perturbation, scaled by inverse square root of the initial local
density $\rho_b$, for the strongly-curved, strong magnetic field (Case C).}
\label{td:f6}
\end{figure}

For this case we measured the oscillatory power in pressure perturbation, scaled
by inverse square root of the initial local density $\rho_b$, which is presented 
in Fig.~\ref{td:f6}. The same straight line structure, as in Cases A and B, is observed.

\begin{figure}
\includegraphics[width=1.0\linewidth]{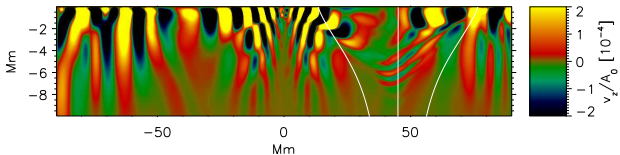}
\caption{Same as Fig.~\ref{td:vx1}, for the strongly-curved strong magnetic field, case C. The slow
mode is visible between the overplotted magnetic field lines.}
\label{td:vx2}
\end{figure}

As in the previous case (Fig.~\ref{td:vx1}), despite the plasma $\beta=1$ surface does 
not appear in the domain,
the slow magnetoacoustic mode of roughly the same amplitude, as for the weakly-curved
magnetic field, is observed in the horizontal velocity component (Fig.~\ref{td:vx2}).
This surprising fact suggests that the conversion of acoustic waves into slow magneto-acoustic 
ones is nearly as efficient as in the case, where the plasma $\beta=1$ surface is located
in the domain. However, here the amplitude of horizontal velocity component in the magnetic 
region is larger, than in the case of weakly-curved strong magnetic field.

\subsection{Overall comparison of the effect of magnetic field structures on wave propagation}
\label{ssc4}
In the previous sections we have shown that a magnetic field of the same strength acts
differently on acoustic waves depending on the geometry and curvature of the field. Here we summarise 
the findings by showing the travel time difference and wave packet amplitude dependencies 
for the representative three analysed cases.

\begin{figure}
\includegraphics[width=1.0\linewidth]{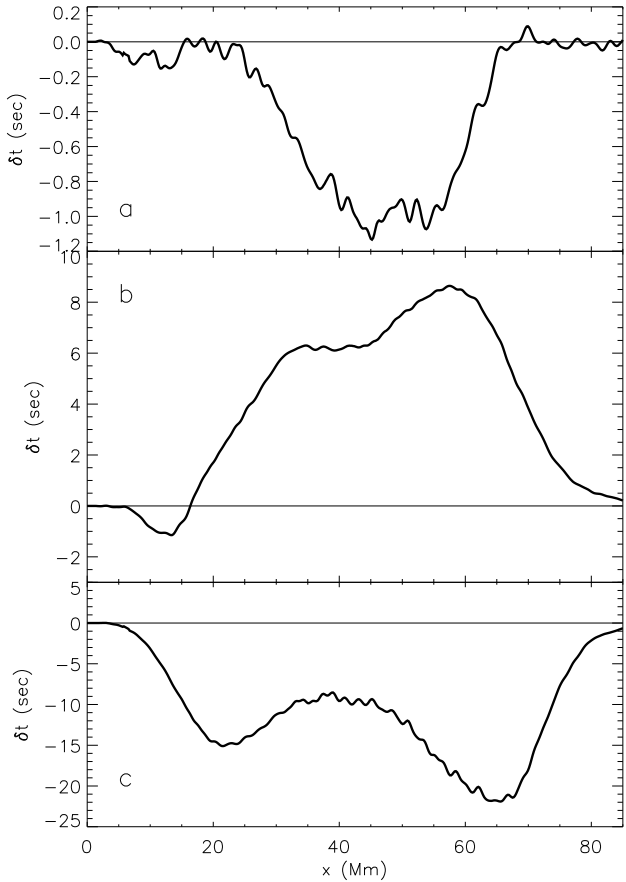}
\caption{Travel time difference plots computed for the first bounce for weak (case A), weakly 
curved strong (case B) and strongly curved strong (case C) magnetic field cases.}
\label{td:f7}
\end{figure}

We have measured the travel time perturbations for the generated waves when travelling
through the flux tube using both G{\'a}bor wavelet and the Gizon-Birch definitions. A similar 
technique was applied by \citet{thompsonzharkov}.
The so-called Gizon-Birch travel time difference plots are shown in Fig.~\ref{td:f7}
for all three magnetic field cases, respectively. Each plot shows that the sign of travel 
time difference, calculated for the first bounce, changes correspondingly to the sign of 
the temperature difference caused by the magnetic field curvature. The travel time differences,
obtained for the simulation with the weakly-curved strong magnetic field (Case B), are of the same order of
those obtained from the observations \citep{duvall, hughes, zharkov07}. This fact suggests that the
magnetic configuration, used for this simulation, is close to the magnetic field structures
in the real sunspots.

\begin{figure}
\includegraphics[width=1.0\linewidth]{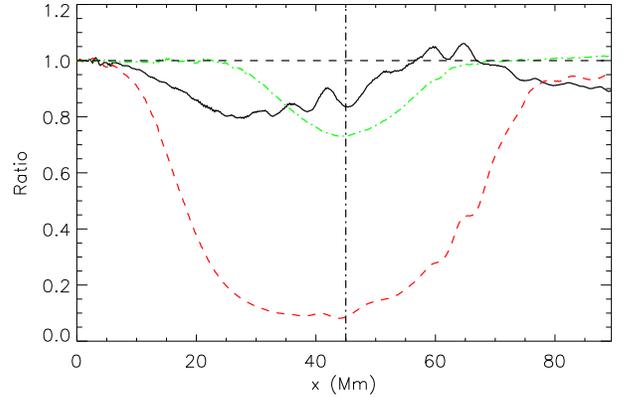}
\caption{Acoustic power absorption by the simulated sunspots. The green dash-dotted, black 
solid and red dashed curves correspond
to the case A, case B and case C magnetic field configurations, respectively.
Horizontal dashed line represents the power ratio 1. Vertical dash-dotted line shows the axis of the 
magnetic configuration.}
\label{td:f10}
\end{figure}

Fig.~\ref{td:f10} demonstrates the difference in the acoustic power absorption and suppression by the
three different magnetic field configurations at the solar surface. The green curve corresponds 
to Case A. The oscillatory power suppression in the magnetic field region 
reaches 30\%, however, no energy absorption is observed, since the power ratio at the distance 
$x=80~\mathrm{Mm}$ is close to unity. This fact, together with the absence of noticeable slow 
magnetoacoustic mode in the horizontal component of velocity, confirms our suggestion that 
the weak magnetic fields act only as temperature and sound speed perturbations for the waves 
propagting through the fields.

The power ratio for Case C (Fig.~\ref{td:f10}, red curve) is
rather different. A very strong suppression of plasma motions in the magnetic field region 
(up to 90\%) is observed. This suppression is caused by the strongly increased temperature in the
simulated sunspot. Also, the energy absorption of about 5\% is obtained at 
$x=80~\mathrm{Mm}$. This confirms the partial conversion of the wave packet energy into slow 
magnetoacoustic mode, which propagates downwards and removes the energy from the solar surface.

An even more complicated behavior is demonstrated by the acoustic power ratio at the solar surface, 
calculated for Case B (Fig.~\ref{td:f10}, black curve). 
In this case, the absorption reaches the value of about 10\% at $x=80~\mathrm{Mm}$. 
The character of energy suppression in the magnetic field region is 
also different. The first feature here is that the curve is not symmetric with respect
to the vertical axis of the magnetic field configuration (note that the green curve, corresponding
to the weak magnetic field (Case A) is completely symmetric around the axis, and the red curve is
also nearly symmetric, if the energy absorption is not taken into account). We suggest that the 
wavy structure between $x=35~\mathrm{Mm}$ and $x=50~\mathrm{Mm}$ is connected to the conversion
of purely acoustic wave into slow magnetoacoustic mode. However, the most noticeable feature
of this curve is the acoustic power increase of the order of 7-8\% at the distance $x=60~\mathrm{Mm}$.
This feature is also clearly visible in the two-dimensional power image (Fig.~\ref{td:f4}), and can be
compared with acoustic power haloes around sunspots, usually observed (see, for example, recent 
observations by \citet{hill} or \citet{nagashima} using new instrument Hinode, and references therein).

The straight lines emanating at different angles from the flux tubes' left
boundary to the right edge of the box, visible in both oscillatory power plots for vertical velocity
component, kinetic energy and pressure perturbation, in our view can be explained in terms of the ray
theory as the caustic surface changes occurring due to the sound-speed
inhomogeneity in the $x$-direction. There are caustics corresponding to the envelope to the
ray paths for the second and higer-order bounces: these are approximately co-located
with the loci of the lower turning points for the second and higer-order bounces,
illustrated in Figs.~\ref{td:f2}, \ref{td:f4}, \ref{td:f6}.

Viewed as the focusing points for the generated
waves, the caustics can be characterised by an increase of the power in the
oscillations \citep{kravorl}. Thus, the ratio of the power between the two cases can be
expected to be most pronounced at such surfaces. 
This power increase is clearly observed in the time series (movies) of the 
simulated  wave field, available in online material at 
http://robertus.staff.shef.ac.uk/publications/acoustic/ .
The magneto-acoustic mode generation and propagation is also clearly
visible in the movies.

\section{Conclusions}

In this paper we presented the numerical modelling and helioseismological analysis 
of three physically different, localised magnetic field concentrations, mimicking 
sunspots in the solar photosphere. The model photosphere is based on the solar Standard 
Model S. The acoustic response of this quiet (non-magnetic) Sun model is close
to that of the real Sun. The implemented magnetic fields for the simulations are
different not only by their strength, but also by the curvature
of the field lines. The curvature of magnetic field creates magnetic tension,
which consequently changes the pressure, density and temperature stratification
of the equilibrium model. Three representative cases of magnetic fields in equilibrium
with the external non-magnetic photospheric plasma are considered: weak magnetic field, strong
but weakly-curved magnetic field and strongly-curved but strong magnetic field models.
As a result of different magnetic field geometry, different temperature structures
were obtained. In the case of weak magnetic field (Case A), the temperature deviation from
the background is small, however, there is a complex structure of temperature decrease
in the photosphere and temperature increase in the sub-photosphere. The two strong 
magnetic field cases (B and C) have the same magnetic field strength at the surface 
($3.5~\mathrm{kG}$). The case of the weakly-curved field (Case B) is characterised by
the temperature decrease below the solar surface. However, the strongly-curved
magnetic field (Case C) makes the temperature increased there.

The spatial structure of the models we used to carry out the simulations is such 
that it allows direct and easy comparison of the behaviour of the waves going 
through the non-magnetic plasma with the behaviour of the waves 
interacting with the magnetic field region. For that, we imposed magnetic 
field only in one half of the numerical domain, leaving the other half unaffected 
by magnetic field. 

We analysed the three magnetic field cases by the means of local time-distance 
helioseismology. Synthetic time-distance, time-distance difference and travel time difference 
dependencies were calculated from the simulations. The dependencies show that the
main part of effect of magnetic field on the acoustic wave is due to the change
of the temperature structure in the sunspot. However, we also show that there is
an energy leakage downwards in the model due to the wave mode conversion from 
purely acoustic to slow magneto-acoustic wave motion. 

Despite the fact that the results are intrinsically correct up to the order of numerical
noise amplitude, we acknowledge the disadvantage of their somewhat limited applicability. 
The simulations are carried out for a magnetic field and background model, which 
are essentially two-dimensional. Thus, the main applicability limitation of our results 
consists in the energy distribution in the acoustic modes which is quantitatively (but 
not qualitatively, if only acoustic and magneto-acoustic waves are considered) different 
from the three-dimensional case. Also, the absorption of 
acoustic waves by a magnetic region in reality may be different from the one presented
due to the difference in the acoustic energy distribution. 
However, since the sound and Alfv{\'e}n speeds, and the other main magnetohydrodynamic
parameters are not affected in any way by the dimensionality of the problem, the travel 
times and travel time differences are also independent on the dimensionality.  
 
The two-dimensional magnetic fields used in the simulations presented in this paper
can be extended to the three-dimensional cylindrically symmetric fields. However, 
simulations of acoustic wave propagation through three-dimensional magnetic structures
requires significantly larger computing resources, so we leave such analysis for the nearest future.

\section{Acknowledgments}
This work was supported by a grant from the UK Science and Technology Facilities 
Council (STFC). RE acknowledges M. K{\'e}ray for patient encouragement. RE is also grateful 
to NSF, Hungary (OTKA, Ref.No. K67746). S. Zharkov acknowledges the support of the HELAS 
European Network.

\end{document}